\documentclass[preprint,
amssymb,floatfix,floats,aps,preprintnumbers,noshowpacs,
]{
revtex4-1}

\usepackage{rcs}
\usepackage{color,graphicx}
\usepackage[dvipsnames]{xcolor}
\usepackage{dcolumn}
\usepackage{upgreek}
\usepackage{bm}
\usepackage{chngcntr}
\usepackage[thinspace,thinqspace]{SIunits}
\usepackage{natbib} 
\usepackage[ulem=normalem]{changes}

\newcommand{\yab}{YbAlB$_{4}$}
\newcommand{\ayab}{$\alpha$-YbAlB$_{4}$}
\newcommand{\byab}{$\beta$-YbAlB$_{4}$}

\newcommand{\To}{$T_{0}$}
\newcommand{\Too}{$T^*$}
\newcommand{\Bl}{$B_{1}$}
\newcommand{\Bu}{$B_{2}$}

\newcommand{\kk}{\mathbf{k}}

\newcommand{\beq}{\begin{equation}}
\newcommand{\eeq}{\end{equation}}

\definechangesauthor[color=ForestGreen]{EOF}
\definechangesauthor[color=Purple]{MSG}

\begin{document} 
\title{Quantum critical behaviour and Lifshitz transition in
intermediate valence \ayab} 

\author{Mihael S. Grbi\'{c},$^{\ast1,2,\dag}$ Eoin C. T. O'Farrell,$^{\ast1,\dag}$ Yosuke Matsumoto,$^{\ast1}$ Kentaro Kuga,$^{1}$ Manuel Brando,$^{3}$ Robert K{\"u}chler,$^{3}$ Andriy H. Nevidomskyy,$^{4}$ Makoto Yoshida,$^{1}$ Toshiro Sakakibara,$^{1}$ Yohei Kono,$^{1}$ Yasuyuki Shimura,$^{1}$ Michael L. Sutherland,$^{5}$ Masashi Takigawa,$^{1,\dag}$ Satoru Nakatsuji$^{1,\dag}$} 
\affiliation{$^{1}$Institute for Solid State Physics (ISSP), University of
Tokyo, Kashiwa 277-8581, Japan}
\affiliation{$^{2}$Department of Physics, Faculty of Science, University
	of Zagreb, Bijeni\v {c}ka 32, Zagreb HR 10000, Croatia}
\affiliation{$^{3}$Max Planck Institute for Chemical Physics of Solids,
	N{\"o}thnitzer Strasse 40, D-01187 Dresden, Germany}
\affiliation{$^{4}$Department of Physics and Astronomy, Rice University,
	Houston, Texas 77005, USA}
\affiliation{$^{5}$Cavendish Laboratory, University of Cambridge, J.J.
	Thomson Avenue, CB3 0HE\vspace{1cm}, UK}
\affiliation{$^{\ast}$These authors contributed equally to this work,}
\affiliation{$^\dag$Correspondence should be sent to: eoin.ofarrell@gmail.com (E.C.T.O.), mgrbic@phy.hr (M.S.G.), masashi@issp.u-tokyo.ac.jp~(M.T.), satoru@issp.u-tokyo.ac.jp (S.N.)}

\date{\today} 
\maketitle

\textbf{ Intermetallic compounds containing $f$-electron elements have been
prototypical materials  for investigating strong electron correlations and quantum criticality (QC). Their heavy fermion ground state evoked by the  magnetic \mbox{$f$-electrons} is susceptible to the onset of quantum phases, such as magnetism or superconductivity, due to the enhanced effective mass ($m^{*}$) and a corresponding decrease of the Fermi temperature. However, the presence of $f$-electron valence fluctuations to a  non-magnetic state is regarded an anathema to QC, as it usually generates a paramagnetic Fermi-liquid state with quasiparticles of moderate $m^{*}$. Such systems are typically isotropic, with a characteristic energy scale $T_0$ of the order of hundreds of kelvins that require large magnetic fields or pressures to promote a valence or magnetic instability.  Here we show that the intermediate valence compound \mbox{\ayab} surprisingly exhibits both quantum critical behaviour and a Lifshitz transition under low magnetic field, which is attributed to the anisotropy of the hybridization between the conduction and localized $f$-electrons. 
These findings suggest a new route to bypass the large valence energy scale in developing the QC.  } 

\section{Introduction}

A quantum critical point (QCP) occurs when the ground state of a system is
continuously tuned between two states\cite{Coleman2005,Sachdev2011}. The strong incipient quantum fluctuations modify the system's electronic state over large regions of its phase diagram. This has led to the notion that
understanding quantum criticality (QC) is the key to understanding emergent phases in materials such as spin liquids and the high temperature superconductors.

Heavy fermion (HF) materials, often $f$-electron based intermetallics containing Ce or Yb, have been prototypical for the investigation of QC\cite{Coleman2005,Sachdev2011,Gegenwart2015,Wirth2016}: the enhanced entropy of the Fermi sea makes them susceptible to low temperature instabilities such as magnetism and superconductivity. 
The energy differences between these possible ground states are small, and therefore can typically be tuned by applying small magnetic fields or pressures.

The conventional paradigm of QC in HF materials, the Doniach phase diagram and its recent extensions\cite{custers2003break}, requires a stable valence of the magnetic ion, which for Yb is the $3+$ state. Yb intermetallics with fluctuating valence, such as YbAl$_3$ or YbAgCu, have a large valence fluctuation energy scale $T_0$ making new phases difficult to achieve; YbAl$_3$ is not known to order magnetically at all \cite{braithwaite2013p}, while YbAgCu$_4$ requires extremely high pressure\cite{Graf1995}.
The recently synthesized \yab\ is therefore remarkable because QC coexists with intermediate valence, $+2.73$ and $2.75$ for $\alpha$ and $\beta$ polymorphs \cite{Okawa2010}, respectively. In this article we focus on \ayab\ and show the presence of QC that is tuned with small magnetic fields. 

The two polymorphs of \yab\ are locally isostructural, with a highly anisotropic structure where sheets of boron separate the layers containing Yb and Al. The local structure is shown in Fig.~\ref{fig1}a) illustrating the anisotropy, atypical for intermediate valence (IV) systems, and the slightly lowered symmetry of \ayab. Both materials have a valence fluctuation scale $T_0\approx200$~K \cite{Okawa2010, Matsumoto2011} and only a minor change of valence~\cite{Terashima2015} in magnetic fields up to 40\,T.

\section{Summary of main results} We use a complete set of thermodynamic, magnetotransport and microscopic experimental techniques to probe the
electronic anisotropy as \ayab\ is driven toward two electronic instabilities with a magnetic field
that is small compared to \To$\approx 200~$K.  Most prominently, we find that the sign of the
thermal expansion, which directly probes the pressure dependence of the entropy \cite{Garst2005}, changes at
$B_\mathrm{c}=3.6$~T due to a change in the nature of the relevant fluctuation scale from
magnetic correlations at lower fields, to the Kondo or valence correlations at higher fields; a signature of
the proximity to a QCP. At a slightly lower field $B_\mathrm{c}=2.1$~T, the Shubnikov--de Haas
measurements show the appearance of a new, stri\-kin\-gly anisotropic Fermi surface (FS), indicating a
Lifshitz transition.  We refer to these fields as \Bl\ = 2.1~T and \Bu\ = 3.6 T in what follows.
At both of these fields nuclear magnetic resonance (NMR) measurements find a diverging spin lattice
relaxation rate $1/T_{1}$ of $^{11}$B nuclei down to 50~mK. The magnetostriction, i.e. the rate of
change of the lattice constant with magnetic field, is highly anisotropic: for the $c$-axis the magnetostriction is
maximum at \Bu, while for the $a$-axis it is maximum at \Bl. The resistivity shows non-FL (NFL) behaviour
at \Bl\ and \Bu\, but only for current applied parallel to the $c$-axis. 
 The striking two dimensionality (2D) of the FS that appears at \Bl\ is consistent with
it being a remnant of the Lifshitz transition proposed to drive criticality in \byab \cite{Ramires2012}, while
the change in energy scales at \Bu\ indicates the proximity to a QCP.
We provide a tentative theoretical explanation for this anisotropy in terms of the momentum-dependent nature of the Kondo hybridization, and resolve the mismatch between the large intermediate valence energy scale and the much smaller energy associated with fields \Bl\ and \Bu.

By combining these complementary experimental techniques together with theoretical arguments we establish a complete picture of the system and find that the hybridization anisotropy provides a means to overcome the large valence energy scale in the \yab\ system.

\section{Results} 
\begin{figure*}   
\includegraphics[width=\textwidth]{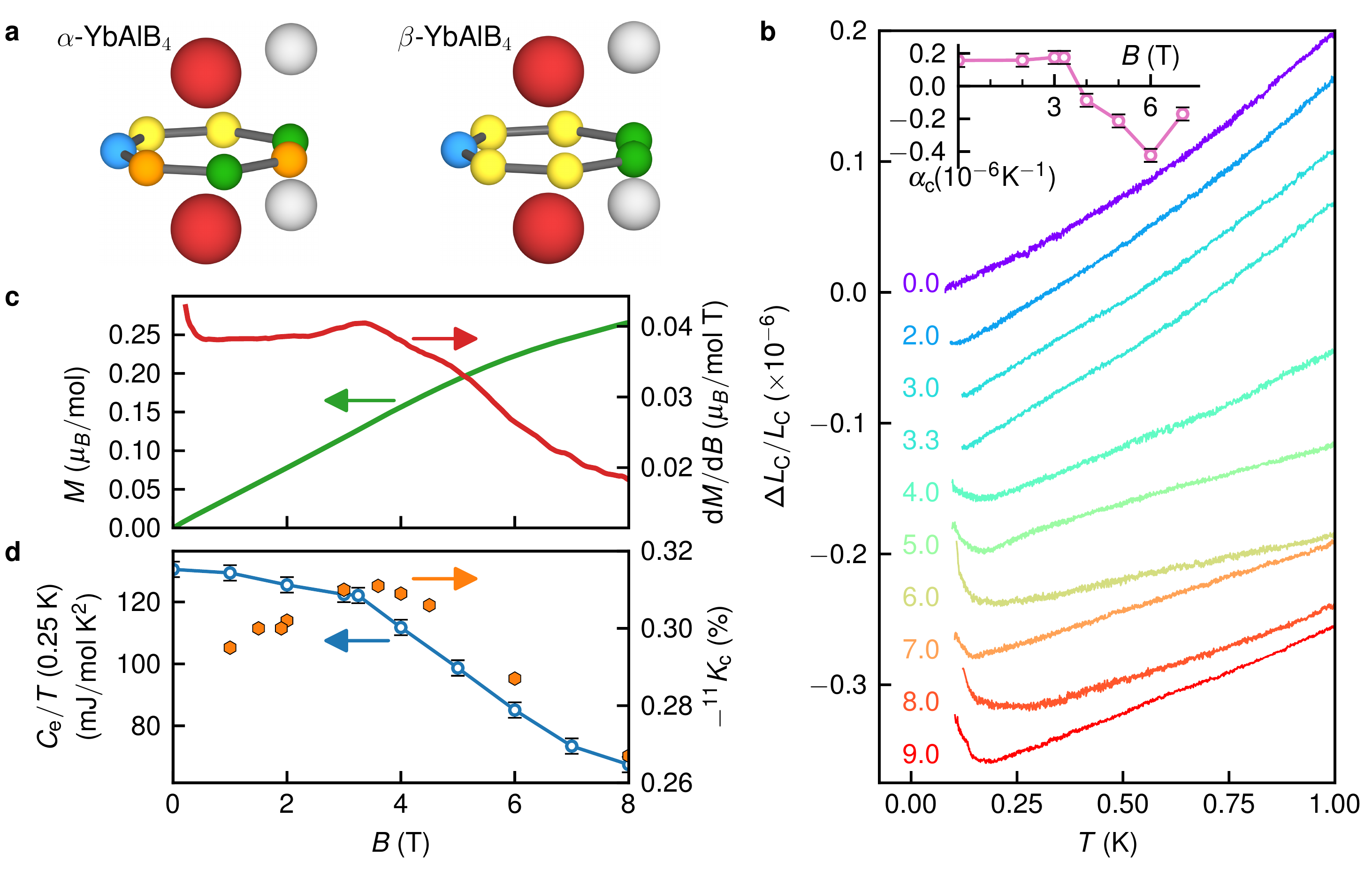}    
\caption{\textbf{Magnetic field  $B \parallel c$-axis behaviour.}     
\textbf{a}, The atomic neighbourhood of Yb ions in \yab\, the Yb site is shown in red together with the surrounding B and Al sites. B sites are coloured in blue, yellow, green and orange according to their symmetry position, while Al is gray.
\textbf{b}, Thermal
expansion $\Delta L_{c}/L_{c}$ and thermal expansion coefficient $\upalpha_{c} = d\left(\Delta L_{c}/L_{c}\right)/dT$ along the $c$-axis for several magnetic fields. The
inset shows the field dependence of $\upalpha_{c}$ taken at 140\,mK.     
\textbf{c}, Magnetization and
susceptibility $\chi = dM/dB$ vs $B \parallel c$ at $T = 0.08$\,K.     
\textbf{d}, Electronic
specific heat coefficient $\gamma_{e}$ measured at 250 mK and Knight shift of the $^{11}$B nucleus vs $B \parallel c$ measured at 142 mK.} 
\label{fig1}
\end{figure*}

We first describe the magnetic field dependence  of the thermal expansion $\Delta
L_{c}/L_{c}$, where $L_{c}$ is the sample length along the $c$-axis, shown in Fig.~\ref{fig1}b. At
$B = 0$ and low temperatures ($T<$ \Too) the linear thermal expansion coefficient $\upalpha_{i}$ is positive for both the $i = a$ and $c$ axes, and thus the volume coefficient
$\upalpha_\mathrm{Vol}>0$.
This is surprising
since in Yb-based Kondo-lattice (KL) or IV systems $\upalpha_\mathrm{Vol}<0$ usually, as is magnetostriction $\lambda_{i} = d\left(\Delta L_{i}/L_{i}\right)/dB$ (see Supplementary Note 1 for details). However, $\upalpha_{Vol}$ measures the
pressure dependence of the entropy, which in \ayab\ implies that the dominant contribution does not
arise from KL or IV type fluctuations, but from an energy scale that increases with pressure. In
magnetic Yb-based systems this is usually (anti-)ferromagnetic (A)FM order
mediated by the RKKY interaction~\cite{Thompson1994,Gegenwart2008}. The high Wilson ratio ($\chi_0 / \gamma_0 \approx 7$) and
$\upalpha_{Vol} (B=0) > 0$ indicates that FM correlations dominate the
ground state. In fact, it has been found that pressure induces an AFM state in \byab\
through a first-order phase transition at
2.5\,GPa~\cite{Tomita2015,Kuga2008}, and an AFM state emerges in \ayab\ by Fe doping\cite{Kuga2012,Kuga2018} at 1.5\%.  Although the AFM critical field of the Fe-doped system~\cite{Kuga2018} is close to \Bu, the features observed at \Bu\ in pure \ayab\ have a different origin and cannot be related to the AFM-to-PM field-induced QCP. We argue this by further discussing our results.

Under magnetic field the magnetic correlations are suppressed and the sign of $\upalpha_{c}$ changes between 3.3\,T and 4\,T, indicating a change
of the relevant energy scale as expected at a crossover or a phase transition between an ordered and
a disordered phase. At a QCP it is expected that $\upalpha_{Vol}/T$ becomes
divergent and changes sign~\cite{Garst2005}. However, in \ayab\ this change is rather
smooth and asymmetric, similar to the one found in YbAgGe~\cite{Gegenwart2016}, and the system remains paramagnetic (PM) on both sides. This behaviour is typical of
metamagnetic materials, like Sr$_{3}$Ru$_{2}$O$_{7}$~\cite{Gegenwart2006} or
CeRu$_{2}$Si$_{2}$~\cite{Weickert2010} where the entropy is dominated by magnetic fluctuations (see Supplementary Note 1 for details). The relatively smooth change in $\upalpha_\mathrm{Vol}/T$ suggests that \ayab\ is located
either close to a quantum critical endpoint, or
in the proximity of a field-induced QCP.

Within both scenarios a clear anomaly in magnetic susceptibility $\chi$ and specific heat coefficient $\gamma$ is expected. The magnetization $M$ and
$\chi = dM/dB$ measured with $B \parallel c$ are shown in Fig.~\ref{fig1} c, with $dM/dB$ displaying
a clear enhancement at \Bu\ before decreasing rapidly at higher magnetic field. Compared with true
metamagnetic materials the enhancement is weak and is not symmetric around \Bu. The same features
are seen in the specific heat coefficient $C_e/T=\gamma$ and in the NMR Knight shift $^{11}K$ measured at the $^{11}B$ nucleus, as
shown in Fig.~\ref{fig1}d. Since all three quantities are governed by the Fermi surface properties: $\gamma \propto \chi \propto\,^{11}K \propto N(\epsilon_{F})$, with
$N(\epsilon_{F})$ the density of states at the Fermi level $\epsilon_{F}$, it is clear that the
anomaly at \Bu\ involves a continuous reduction of $N(\epsilon_{F})$ as a consequence of the
suppression of the correlations by the field. This is different from a possible suppression of the
HF state by magnetic field through a metamagnetic transition. In fact, when compared to HF metamagnetic compounds~\cite{Kitagawa2011} the value of \Bu\ is
not large enough. In
\ayab\ the relevant magnetic field scale, estimated from $\gamma = 130$\,mJ/molK$^{2}$, should exceed
20\,T, and hence metamagnetism cannot account for the observed phenomena.

\begin{figure*}  
\includegraphics[width=\textwidth]{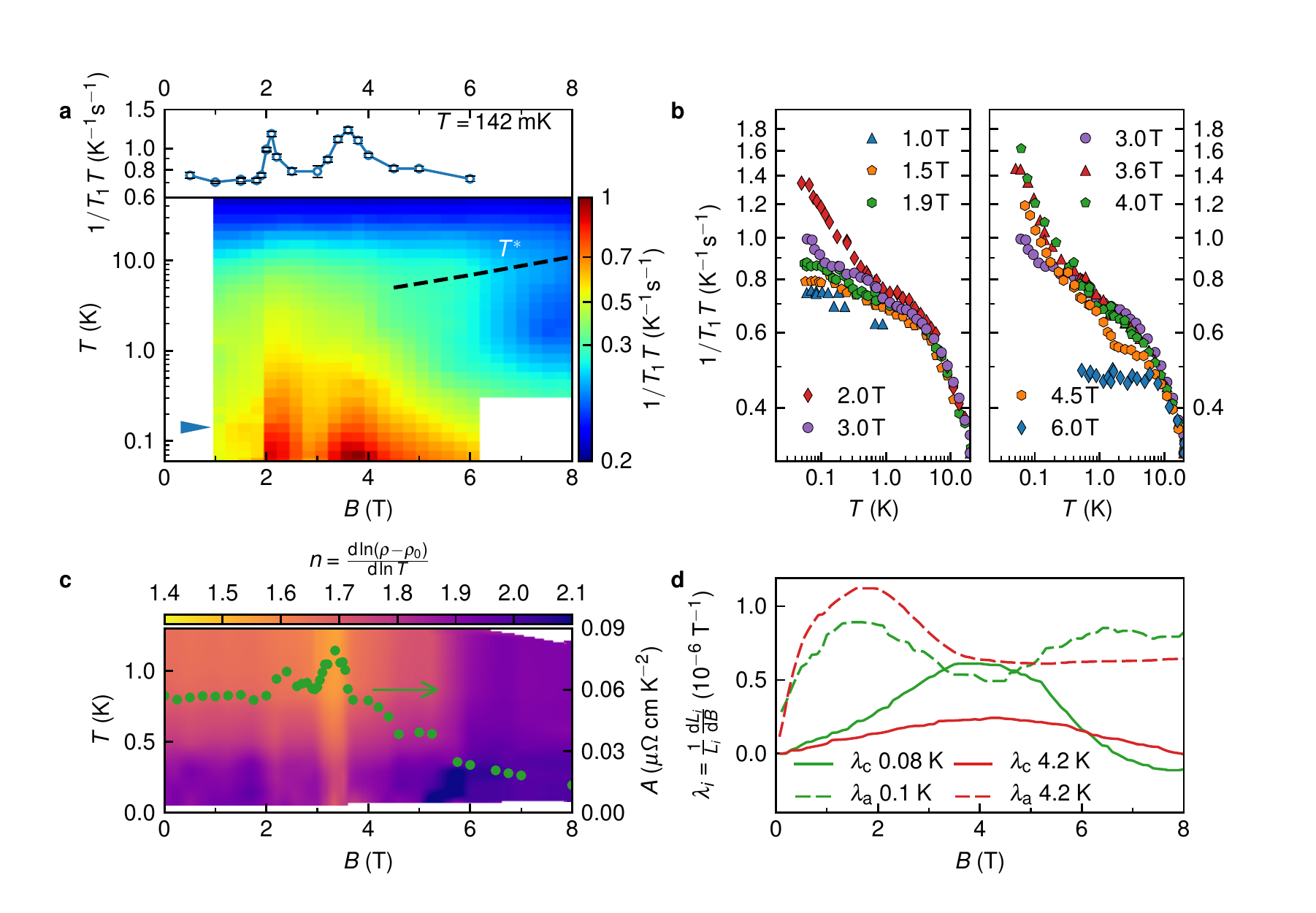}
\caption{\textbf{Detection of quantum criticality by NMR and resistivity.}   
\textbf{a}, (Lower
panel) Density plot forming a $B - T$ phase diagram of $(T_{1}T)^{-1}$ on $^{11}$B   sites showing
the divergent behaviour of spin fluctuations at the two critical fields   \Bl\ = 2.1\,T and \Bu\ =
3.6\,T.   (Upper panel) Cross section of $(T_{1}T)^{-1}$ on $^{11}$B sites at $T=142$ mK as indicated by the arrow in the lower panel.
\textbf{b}, Temperature dependence of $(T_{1}T)^{-1}$ at   fields close to \Bl\ (left panel) and \Bu\ (right panel).   
\textbf{c}, Density plot of the power law exponent of resistivity and the $A$
coefficient extracted at $T = 0.1$\,K vs $B \parallel c$.
\textbf{d}, Magnetostriction coefficient $\lambda$ vs $B$ along both $a$ and $c$-axes at $T$ = 0.1\,K and $T$ =
4.2\,K.}
\label{fig2}
\end{figure*}

A rigorous test of the ground state properties is the character of underlying excitations. To probe them  we have measured the NMR relaxation rate $1/T_{1}$ (see Methods) of the $^{11}B$ nuclei (plotted in Figs.~\ref{fig2}a,b and c) as a function of temperature at various magnetic fields for $B
\parallel c$. The data are shown as $(T_{1}T)^{-1}$ so that they can be easily contrasted to $(T_{1}T)^{-1}$= const. that is typically observed for simple metals at low temperature. Below \Too\ $\approx 8$\,K and below 1.5 T, where the coherent FL state is formed, $(T_{1}T)^{-1}$ is $\approx 0.7$~K$^{-1}$s$^{-1}$. This is two orders of magnitude lower than what is typical with dominant magnetic correlations~\cite{Kohori2001,Kitagawa2011,Kitagawa2013}, which shows that the system is not close to magnetism. As the magnetic field approaches \Bu, we surprisingly find a pronounced divergence $(T_{1}T)^{-1} \propto T^{-\delta}$ at low temperatures with exponent $\delta = 0.36$. This power-law behaviour persists down to 50~mK, two orders of magnitude lower than \Too, and reveals the presence of quantum critical fluctuations that destabilize the FL and lead to a NFL ground state. The exponent $\delta =0.36 < 1$ show that this is not a magnetic QCP - it is too low for the divergence to appear in $1/T_{1}$ alone, which would be usual for a magnetic QCP.
Also, across \Bu\ the NMR spectrum remains unchanged (Supplementary Fig. 1) which would split if magnetic or charge order appeared at \Bu, due to symmetry breaking.

We have also analysed the Knight shift behaviour $K = \langle b_{z} \rangle/B$, where $\langle b_{z}
\rangle$ is the hyperfine magnetic field at the nucleus site. While $(T_{1}T)^{-1}$ is sensitive to spin excitations at $\textbf{q} \geq 0$, the Knight shift is related to the static spin susceptibility $\chi(\omega = 0, \textbf{q} \approx 0)$, and in a FL it is expected to be
constant against temperature and magnetic field. As seen in Fig.~\ref{fig1}d, for $B <$\Bu\ the $c$-axis Knight
shifts of $^{11}$B ($^{11}K_{c}$) and $^{27}$Al ($^{27}K_{c}$) nuclei (Supplementary Fig. 2) are constant, and at \Bu\ show a weak
feature (similar to $\chi$) confirming the absence of magnetic $\textbf{q}=0$ mode at \Bl\
and \Bu. Across 4\,T, both
$^{11}K_{c}$ and $^{27}K_{c}$ show only a slow change arising from $N(\epsilon_{F})$, which is also supported by the measurements of the quadupolar coupling of $^{27}$Al and $^{11}$B (Supplementary Figs. 3 and 4).

To summarize, from the presented data we can deduce that at \Bu\ the underlying ground state is NFL, no magnetism emerges, the critical spin fluctuations originate at $\textbf{q} \neq 0$ and there is a change in the static charge density distribution.

So far, we have focused on the QC at \Bu\ $\approx 3.6$\,T visible in
thermodynamic and NMR measurements. In addition, the NMR measurements show the existence of another
unusual  behaviour that suddenly appears at a lower field \Bl\ $\approx 2.1$\,T.
Indeed, the magnetic
field dependence of the NMR relaxation rate $(T_{1}T)^{-1}$ at 142\,mK shows two pronounced peaks
(Fig.~\ref{fig2}a) at \Bl\ and \Bu. The temperature dependence 
reveals another NFL
power-law divergence at \Bl, $(T_{1}T)^{-1} \propto T^{-\delta}$, with $\delta = 0.25$ similar to
the behaviour at \Bu. 
Between \Bl\ and \Bu\ the system behaves as a standard metal,  down to lowest temperatures, which indicates
$B_{1,2}$ are separate phenomena. Their qualitatively different nature is evidenced by the critical dynamics at \Bl\ that, unlike the crossover at \Bu, shows no distinct
signature in susceptibility or specific heat despite clear evidence of divergent spin fluctuations from the NMR.

Further indications about the nature of the transition at \Bl\ are given by the magnetostriction
coefficient $\lambda$ measured along both the $a$-axis ($\lambda_a$) and $c$-axis ($\lambda_c$) but with magnetic field $B \parallel
c$, shown in Fig.~\ref{fig2}d. $\lambda$, which is sensitive to structural, magnetic and
electronic structure transitions, here shows a strong anisotropy: $\lambda_{c}(B)$ shows a maximum at \Bu\ as expected from
$\upalpha_{c}$, while $\lambda_{a}(B)$ shows a maximum at the smaller field \Bl\ implying that it is related to a 2D effect within the $ab$-plane.

Anisotropy in the critical behaviour of \ayab\ is also evidenced by resistivity measurements along the $a$
and $c$ axes, $\rho_a$ and $\rho_c$, respectively. Low temperature NFL behaviour due to QC is quantified by observing the exponent $n_i$ and the coefficient $A_i \propto m_i^2$ as $\rho_i = \rho_{0,i} + A_{i}T^{n_i}$, $i=a,c$ (see Methods). At $T=0.1$~K, $A_c$ shows an enhancement (Fig.~\ref{fig2}c) by $\approx50\%$ at \Bl\ and \Bu\, before decreasing at higher fields, and the exponent $n_c$ shows deviations (Fig.~\ref{fig2}c) from the FL value of 2 in the vicinity of \Bl\ and \Bu\ -- $n_c (B_2) = 1.65$ at the lowest measured temperature $T = 0.04$\,K. This is not observed\cite{Kuga2018} in the Fe-doped \ayab\ at 3.5 T, where the exponent retains the FL value of 2, and is therefore different from the pristine sample. The $n_c (B_2) = 1.65$ value in the pristine \ayab\ is close to
that expected at a FM or AFM QCP at which $n = 5/3$ and 3/2, respectively~\cite{Loehneysen2007}. For instance, 
\byab\ shows $n=3/2$ at its critical point consistent with the AFM QCP. Importantly, the NFL behaviour in \ayab\ occurs only for
$\rho_c$, whereas $n_a$ remains $\approx 2$ for $\rho_a$ (Supplementary Fig. 5). This shows that quantum critical fluctuations do not originate from the suppression of the AFM phase in Fe-doped case, and that they manifest themselves 
in specific regions of momentum space, which confirms the conclusion derived from anisotropic magnetostriction and the NMR data. Intrinsic electronic anisotropy is therefore an important factor for understanding this system.

\begin{figure*}   
\includegraphics[width=\textwidth]{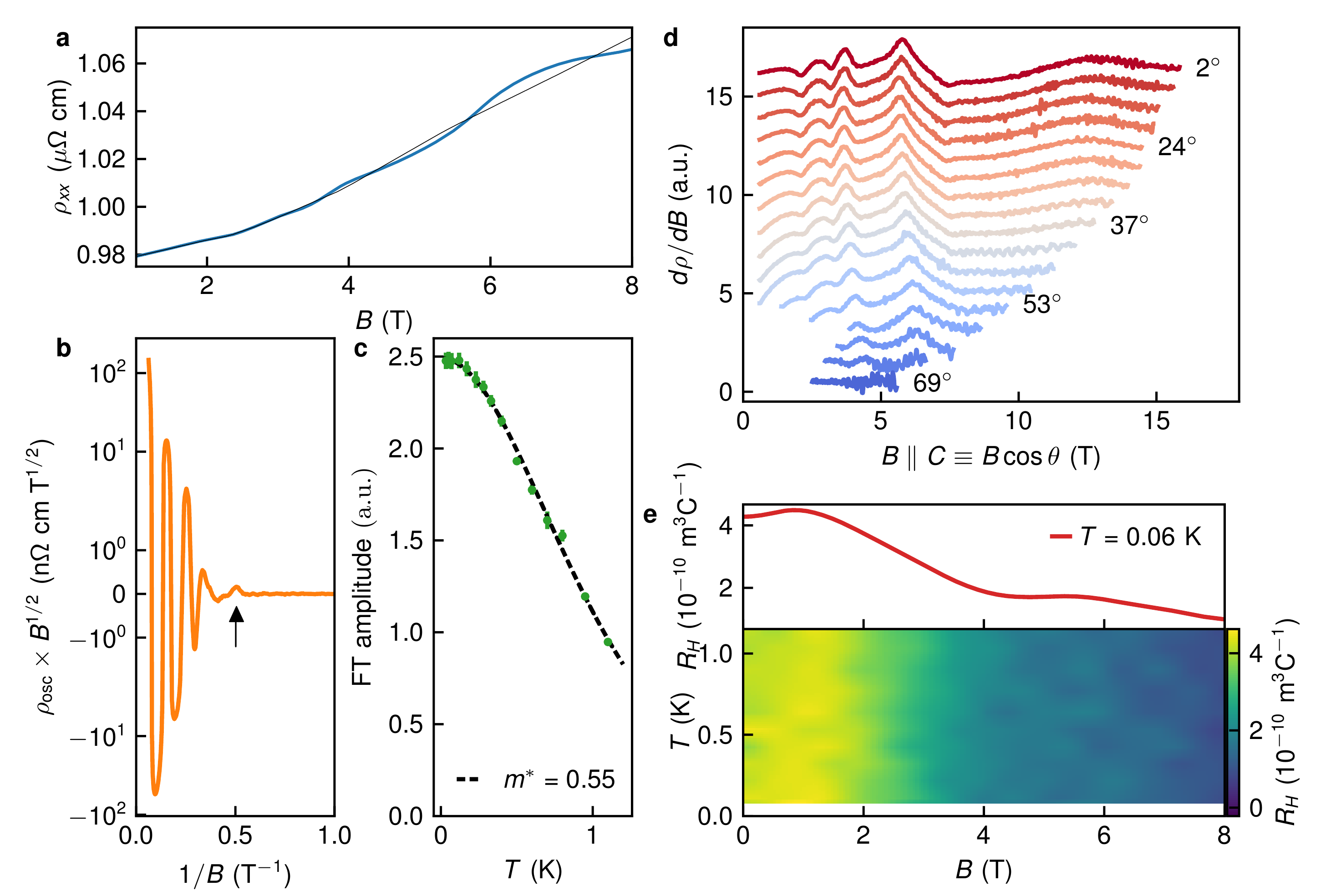}   
\caption{\textbf{Quantum
oscillations and Hall effect in \ayab.}   
\textbf{a}, $\rho,I \parallel c$ vs $B \parallel c$ at $T=0.03$ K, the overlaid black line is the non-oscillatory component, $\rho_\mathrm{MR}$ that is subtracted.  
\textbf{b}, The oscillatory component of resistivity; $\rho_\mathrm{osc}\times B^{1/2}$ plotted on a double logarithmic scale (values $< 10^{-3}$ are plotted on a linear scale). 
\textbf{c}, Temperature dependence of the Fourier transform of $\rho_\mathrm{osc}$.
\textbf{d}, $d\rho/dB$ against $B \parallel c$ at various angles of $B$ in the $[001]\rightarrow[110]$ plane where $\theta=0\equiv B\parallel[001]$.
\textbf{e}, (Upper panel) Hall coefficient $B\parallel$c at $T=~0.06$ K. (Lower panel) surface plot of Hall coefficient against field and temperature.}   
\label{fig3} 
\end{figure*}

We now consider the changes in the electronic structure in the vicinity of \Bl\ and \Bu, by measuring quantum oscillations (QO). Figure~\ref{fig3}(a) shows $\rho_c$ as a function of the magnetic field, from which we separate the slow-varying ($\rho_\mathrm{MR}$) and the oscillatory ($\rho_\mathrm{osc}$) component (see Methods). A clear kink in $\rho_\mathrm{osc}\times B^{1/2}$ (Fig.~\ref{fig3}(b)) is observed at $1/B=0.5$ below which oscillations appear. The logarithmic scale is used to show that the amplitude decays linearly as expected from the Dingle relation. The amplitude is also shown against temperature in Fig.~\ref{fig3}(c) together with a fit to the Lifshitz-Kosevich (LK) relation, that describes the decay of QOs with $T$ and gives $m^*=0.55\pm0.03$. The excellent agreement to the LK relation confirms these are indeed QOs with frequency $F=10.2$ T, while their sudden appearance at \Bl\ suggests they emerge as the result of a Lifshitz transition at \Bl. This remarkable result makes \ayab\ the first IV compound with a Lifshitz transition induced by such a low magnetic field.

By rotating the sample relative to $B$ we probe the extremal cross-section of this FS perpendicular to $B$. Figure~\ref{fig3}(d) shows $d\rho/dB$ against the $c$-axis component of $B$, i.e. $B\cos \theta$ where the $B$ is rotated in the plane of [001] and [110]. The $\rho_\mathrm{osc}$ is observed to be constant with $B\cos \theta$ indicating the FS is cylindrical and arises from 2D carriers, the volume of this cylinder is very small, corresponding to $4\cdot 10^{-4}$ carriers per Brillouin zone. The QOs of different spin components of a 2D FS are normally split by the Zeeman interaction and interfere to produce spin zeros where the QO amplitude vanishes and the phase shifts by $\pi$. Using a $g$ factor of 2.3 obtained by electron spin resonance \cite{Holanda2011} and the determined value of $m^*$, we would expect a strong angular dependence of the
amplitude and a spin zero at $\theta=60^\circ$ (see Supplementary Note 3 for details).  However, no spin zeros are found, suggesting that this pocket is spin polarized, consistent with a Zeeman-driven Lifshitz transition at \Bl. If this pocket is fully spin polarized and taking\cite{Nevidomskyy2009} $J_z=5/2$, the magnetic moment of this pocket is $ 0.002 \mu_B$, which is consistent with the extremely small polarization increase observed in $M$ in Fig.~\ref{fig1}c).

The 2D nature of the FS, the strong divergence in $(T_{1}T)^{-1}$ at the $^{11}$B site and the pronounced maximum in $\lambda_{a}$ indicate that due to anisotropy the electronic properties within boron layers are most affected at \Bl. However, the small size of the FS pocket relative to
the total number of carriers makes it difficult to account for the strong quantum critical fluctuations and
suggests that larger sheets of the FS are affected. While large QO frequencies are observed
at higher magnetic fields (see Supplementary Note 3 for details), these were difficult to observe at low
magnetic fields of \Bl\ and \Bu, and we therefore turn to the Hall coefficient ($R_\mathrm{H}$) shown in
Fig.~\ref{fig3}e. $R_\mathrm{H}$ decreases smoothly above \Bl, which would within a single-band model indicate a change in the carrier density by a factor of 2. However,
this is not supported by the small changes in $\rho$, and therefore it could be interpreted as changes in the FS velocity on larger sheets of the FS.
The $(T_{1}T)^{-1}$ data and the gradual change in $R_\mathrm{H} (H,T)$ confirm that \ayab\ is close to a QCP at the field \Bu
.

\section{Discussion} 

Intermediate valence compounds have been mainly outside the main focus of research on QC since, due to large characteristic energy scales of the valence fluctuations (\To\ $> 400$\,K in YbAl$_{3}$), application of large magnetic fields or pressures is required to notably modify their properties or induce new phases~\cite{Lawrence1981}.
In contrast, for both \yab\ polymorphs magnetic moments survive well below \To\ $\approx 200$\,K, and become strongly
correlated below \Too\ $\approx 8$\,K, showing QC at low magnetic fields. Although the quantum critical behaviour in the two polymorphs has different phe\-no\-me\-no\-lo\-gy, the presence
of highly anisotropic Kondo hybridization in both compounds appears crucial for establishing it. The conventional picture of IV materials contains the hybridization energy scale $\Gamma = \pi N(\epsilon_{F}) |V|^2$, where  $V$ is Anderson's coupling strength between conduction band and localized $f$-electrons. In the strong coupling limit $\Gamma$  
approaches $T_K$. The NFL behaviour and Kondo hybridization can be connected~\cite{Nevidomskyy2009} by appealing to the tensorial, momentum-dependent nature of the hybridization stemming from the dominant $m_J=5/2$ nature of Yb ground state doublet $H_V = \hat{V}_{\sigma\alpha}(\kk) c^{\dagger}_{\kk\sigma}\, f_{\kk\alpha} + \text{h.c.},$
expressed in terms of the creation/annihilation operators of conduction ($c^{\dagger}$) and localized ($f$) electrons.
Subsequent theoretical work on \byab\ showed~\cite{Ramires2012} that the tensor $\hat{V}_{\sigma\alpha}(\kk)$ is indeed highly anisotropic and vanishes upon approaching the $\Gamma-Z$ line in the Brillouin zone as $|\hat{V}_{\kk}| \propto \sin(k_z c) k_\perp^2$, which is also confirmed by recent ARPES~\cite{Bareille2018} measurement.
In the lower symmetry polymorph \ayab, this dispersion is not expected to persist, however, the observation of a Lifshitz transition of the strikingly 2D FS pocket and the deviation from FL behaviour suggest that \Bl\ is remnant of the QCP in \byab.
In this case \Bl\ is detuned from zero field by a non-zero value of the renormalized chemical potential $\epsilon_F^{\ast}$: $\xi(\mathbf{k}) = \epsilon_F^{\ast} - g\mu_B B + N(\epsilon_{F})\, \mathrm{Tr}(|\hat{V}(\mathbf{k})|^2)$, whereas $\epsilon_F^{\ast}$ is believed to be zero~\cite{Ramires2012} in \byab. 
This is consistent with the experimental data in \ayab\, if $\epsilon_F^{\ast}$ is $g \mu_B B_1 \approx 2$~meV, such that the signature of QC appears at $B_1=2.1$~T, when the chemical potential reaches the bottom of the majority-spin band.

Unlike the transition at \Bl, the NFL behaviour at \Bu\ is much more a 3D phenomenon that drastically changes the global properties of the system. The FM fluctuations present in the ground state at zero field (shown by $\upalpha_{Vol} > 0$ and high Wilson ratio) are suppressed as the magnetic field is increased to \Bu.
Near \Bu, although FM ($\textbf{q}\sim0$) fluctuations are suppressed, finite-$\textbf{q}$ fluctuations get enhanced, causing power-law divergence of $(T_1T)^{-1}$ and anomaly in resistivity, susceptibility and specific heat, which imply the proximity to a QCP unrelated to the Fe-doped case. At the same time, magnetostriction, QO and thermal expansion show that the transition is followed by a FS change. 
The QCP also differs from the one in \byab, where the well understood
FS of \byab~\cite{Eoin2009,Eoin2012,Ramires2012,Bareille2018} has no small pocket like the one observed to appear at \Bl~in \ayab, and signatures of spin zeros were observed on larger FS sheets \cite{EoinThesis} which are absent in the \ayab.


Although in IV compounds magnetic field can induce a valence QCP~\cite{Watanabe2010}, our measured data and the band structure analysis exclude this for the case of \ayab\ (see Supplementary Note 2 for details). Recently~\cite{Kobayashi2018}, M\"{o}ssbauer spectroscopy found a change in the quadrupolar moment of Yb between 1 and 5 T. This implies the QC at \Bu\ originates from a multipole-type QCP in \ayab. However, due to the low symmetry of the Yb site, its understanding will require further measurements which are beyond the scope of the current work.

It is remarkable that critical fluctuations survive in $\alpha$- and \byab, which are both IV materials -- particularly since there has been only one reported quantum critical IV compound so far: CeRhSn~\cite{Tokiwa2015}. Although it is also anisotropic, there the 2D frustrations drive the QC. This is indicative of the fact that in YbAlB$_{4}$ family a robust physical mechanism emerges in the presence of strong electron anisotropy, that is unaffected by the change of local symmetry. What makes \yab\ family special are the Yb chains interpenetrating~\cite{MatsumotoScience} the sheets of B, which cause the anisotropic hybridization and allow for QC to emerge. The tensorial nature of the Kondo hybridization and its vanishing at certain high-symmetry points in the Brillouin zone is also a central thesis of the theory of topological Kondo insulators~\cite{Dzero2010}, and has shown promising results in a recent theoretical treatment of the quasi-1D Kondo lattice~\cite{Komijani2018}. However, a more detailed analysis is required to encompass the richness of the phase diagram of \yab\ family. 

From all the data shown above we can conclude that anisotropy, along with magnetic frustration, offer a new route of overcoming the high energy scale of IV compounds resulting in quantum criticality and new phases of matter.

\section*{Materials and Methods}
Single crystals of \ayab\ were grown from Al flux. The stoichiometric ratio of Yb:4B was heated in excess Al in an alumina crucible under an Ar atmosphere as described elsewhere~\cite{Macaluso}. Chemical compositions of single crystals were determined by a inductively coupled plasma - atomic emission spectrometry (ICP-AES, HORIBA JY138KH ULTRACE) at ISSP, and the analysis of both polymorphs are in good agreement with the ideal compositions of YbAlB$_4$ within the error bars. We analysed diffraction patterns to determine the crystal structure and lattice constant using the Rietveld
analysis program PDXL (Rigaku) and found no impurity phase. The thermal expansion and magnetostriction were measured with a high-resolution capacitive CuBe dilatometer in a dilution refrigerator~\cite{MagnetostrictionRSI}. Specific heat was measured using relaxation calorimetry~\cite{CalorimetryRSI}. Further detail and subtraction of nuclear contributions will be described elsewhere. Magnetization was measured using capacitance Faraday method~\cite{MagnetizationM}. \\
\indent Resistivity and Shubnikov-de Haas measurements were performed using conventional lock-in amplifier techniques. For measurements with $I\parallel c$ pristine crystals were used, while for $I\perp c$ and for Hall effect measurements larger crystals were polished to form thin plates perpendicular to the $c$-axis.
Low temperature deviations from Fermi liquid (FL) behaviour due to quantum critical fluctuations are quantified by expressing $\rho_i = \rho_{0,i} + A_{i}T^{n_i}$. Taking $n=2$ we extracted $A_i(T)\propto m_i^2$. Similarly, we extracted the temperature exponent by assuming $A$ is constant in $T$ as $n_c=d\ln\delta\rho_c/d\ln T$.\\
\indent For quantum oscillations the small number of oscillation periods made the extraction of the oscillatory component of the resistance challenging; we applied two methods that gave consistent results. We assume that $\rho = \rho_\mathrm{MR}+\rho_\mathrm{osc}$ where $\rho_\mathrm{MR}$ is assumed to be slowly varying. In the first method we apply a low pass filter in $1/B$ that subtracts a locally quadratic polynomial, this is described in detail elsewhere \cite{OFarrell_IOP_2012}. In the second we first subtract a linear component from the entire field range and then take a derivative. A comparison between the Fourier spectrum of these methods is shown in the supporting information (Supplementary Note 3).\\
\indent The NMR measurements of $^{11}$B were performed using a pulsed spectrometer. The spectra were collected by Hahn echo sequence $\pi/2 - \tau -\pi$ with typical value of $\tau = 100~\mu$s, and a $\pi/2$ pulse of 6 $\mu$s. The $T_1$ measurements were performed using a saturation-recovery technique on a satellite NMR line determined in previous work\onlinecite{Takano2016}. Data were fitted to a magnetic relaxation function of the form: $M(t)=M_0 (1-(1/10) e^{-(t/T_1)^b}- (5/10) e^{-(3t/T_1)^b}-(4/10) e^{-(6t/T_1)^b})$.  The sample was oriented in situ by observing the quadrupolar splitting of $^{27} Al$ using a two-axis goniometer for measurement temperatures 1.5 - 300 K, and with a single axis goniometer for measurements at lower temperatures. In all cases the magnetic field orientation was within 2$^{\circ}$ of the crystal c axis.



\bibliography{BibList}{} 

\begin{thebibliography}{10}
\expandafter\ifx\csname url\endcsname\relax
  \def\url#1{\texttt{#1}}\fi
\expandafter\ifx\csname urlprefix\endcsname\relax\def\urlprefix{URL }\fi
\providecommand{\bibinfo}[2]{#2}
\providecommand{\eprint}[2][]{\url{#2}}

\bibitem{Coleman2005}
\bibinfo{author}{Coleman, P.} \& \bibinfo{author}{Schofield, A.~J.}
\newblock \bibinfo{title}{Quantum criticality}.
\newblock \emph{\bibinfo{journal}{Nature}} \textbf{\bibinfo{volume}{433}},
  \bibinfo{pages}{226} (\bibinfo{year}{2005}).

\bibitem{Sachdev2011}
\bibinfo{author}{Sachdev, S.} \& \bibinfo{author}{Keimer, B.}
\newblock \bibinfo{title}{Quantum criticality}.
\newblock \emph{\bibinfo{journal}{Physics Today}}
  \textbf{\bibinfo{volume}{64}}, \bibinfo{pages}{29} (\bibinfo{year}{2011}).

\bibitem{Gegenwart2015}
\bibinfo{author}{Gegenwart, P.}, \bibinfo{author}{Steglich, F.},
  \bibinfo{author}{Geibel, C.} \& \bibinfo{author}{Brando, M.}
\newblock \bibinfo{title}{Novel types of quantum criticality in heavy-fermion
  systems}.
\newblock \emph{\bibinfo{journal}{Eur. Phys. J. Special Topics}}
  \textbf{\bibinfo{volume}{224}}, \bibinfo{pages}{975} (\bibinfo{year}{2015}).

\bibitem{Wirth2016}
\bibinfo{author}{Wirth, S.} \& \bibinfo{author}{Steglich, F.}
\newblock \bibinfo{title}{Exploring heavy fermions from macroscopic to
  microscopic length scales}.
\newblock \emph{\bibinfo{journal}{Nature Reviews Materials}}
  \textbf{\bibinfo{volume}{1}}, \bibinfo{pages}{16051} (\bibinfo{year}{2016}).

\bibitem{custers2003break}
\bibinfo{author}{Custers, J.} \emph{et~al.}
\newblock \bibinfo{title}{The break-up of heavy electrons at a quantum critical
  point}.
\newblock \emph{\bibinfo{journal}{Nature}} \textbf{\bibinfo{volume}{424}},
  \bibinfo{pages}{524--527} (\bibinfo{year}{2003}).

\bibitem{braithwaite2013p}
\bibinfo{author}{Braithwaite, D.} \emph{et~al.}
\newblock \bibinfo{title}{(p, {T}, {H}) phase diagram of heavy fermion systems:
  Some systematics and some surprises from {Ytterbium}}.
\newblock \emph{\bibinfo{journal}{Journal of Superconductivity and Novel
  Magnetism}} \textbf{\bibinfo{volume}{5}}, \bibinfo{pages}{1775--1780}
  (\bibinfo{year}{2013}).

\bibitem{Graf1995}
\bibinfo{author}{Graf, T.}, \bibinfo{author}{Movshovich, R.},
  \bibinfo{author}{Thompson, J.~D.}, \bibinfo{author}{Fisk, Z.} \&
  \bibinfo{author}{Canfield, P.~C.}
\newblock \bibinfo{title}{Properties of $\mathrm{YbAgCu}_4$ at high pressures
  and magnetic fields}.
\newblock \emph{\bibinfo{journal}{Phys. Rev. B}} \textbf{\bibinfo{volume}{52}},
  \bibinfo{pages}{3099} (\bibinfo{year}{1995}).

\bibitem{Okawa2010}
\bibinfo{author}{Okawa, M.} \emph{et~al.}
\newblock \bibinfo{title}{Strong valence fluctuation in the quantum critical
  heavy fermion superconductor
  $\ensuremath{\beta}\mathrm{\text{-}}\mathrm{YbAlB}_{4}$: A hard x-ray
  photoemission study}.
\newblock \emph{\bibinfo{journal}{Phys. Rev. Lett.}}
  \textbf{\bibinfo{volume}{104}}, \bibinfo{pages}{247201}
  (\bibinfo{year}{2010}).

\bibitem{Matsumoto2011}
\bibinfo{author}{Matsumoto, Y.}, \bibinfo{author}{Kuga, K.},
  \bibinfo{author}{Tomita, T.}, \bibinfo{author}{Karaki, Y.} \&
  \bibinfo{author}{Nakatsuji, S.}
\newblock \bibinfo{title}{Anisotropic heavy-{Fermi}-liquid formation in
  valence-fluctuating $\ensuremath{\alpha}$-$\mathrm{YbAlB}_{4}$}.
\newblock \emph{\bibinfo{journal}{Phys. Rev. B}} \textbf{\bibinfo{volume}{84}},
  \bibinfo{pages}{125126} (\bibinfo{year}{2011}).

\bibitem{Terashima2015}
\bibinfo{author}{Terashima, T.~T.} \emph{et~al.}
\newblock \bibinfo{title}{X-ray absorption spectroscopy in the heavy fermion
  compound $\alpha$-$\mathrm{YbAlB}_4$ at high magnetic fields}.
\newblock \emph{\bibinfo{journal}{Journal of the Physical Society of Japan}}
  \textbf{\bibinfo{volume}{84}}, \bibinfo{pages}{114715}
  (\bibinfo{year}{2015}).

\bibitem{Garst2005}
\bibinfo{author}{Garst, M.} \& \bibinfo{author}{Rosch, A.}
\newblock \bibinfo{title}{Sign change of the gr\"uneisen parameter and
  magnetocaloric effect near quantum critical points}.
\newblock \emph{\bibinfo{journal}{Phys. Rev. B}} \textbf{\bibinfo{volume}{72}},
  \bibinfo{pages}{205129} (\bibinfo{year}{2005}).

\bibitem{Ramires2012}
\bibinfo{author}{Ramires, A.}, \bibinfo{author}{Coleman, P.},
  \bibinfo{author}{Nevidomskyy, A.~H.} \& \bibinfo{author}{Tsvelik, A.~M.}
\newblock
  \bibinfo{title}{$\ensuremath{\beta}\mathrm{\text{-}}\mathrm{YbAlB}_{4}$: A
  critical nodal metal}.
\newblock \emph{\bibinfo{journal}{Phys. Rev. Lett.}}
  \textbf{\bibinfo{volume}{109}}, \bibinfo{pages}{176404}
  (\bibinfo{year}{2012}).

\bibitem{Thompson1994}
\bibinfo{author}{Thompson, J.~D.} \& \bibinfo{author}{Lawrence, J.~M.}
\newblock \bibinfo{title}{High pressure studies --- physical properties of
  anomalous {Ce}, {Yb} and {U} compounds}.
\newblock In \bibinfo{editor}{K.~A.~Gschneidner, J.}, \bibinfo{editor}{Eyring,
  L.}, \bibinfo{editor}{Lander, G.~H.} \& \bibinfo{editor}{Choppin, G.~R.}
  (eds.) \emph{\bibinfo{booktitle}{Handbook on the Physics and Chemistry of
  Rare Earths}}, vol.~\bibinfo{volume}{19}, \bibinfo{pages}{383}
  (\bibinfo{publisher}{North-Holland, Amsterdam}, \bibinfo{year}{1994}).

\bibitem{Gegenwart2008}
\bibinfo{author}{Gegenwart, P.}, \bibinfo{author}{Si, Q.} \&
  \bibinfo{author}{Steglich, F.}
\newblock \bibinfo{title}{Quantum criticality in heavy-fermion metals}.
\newblock \emph{\bibinfo{journal}{Nature Phys.}} \textbf{\bibinfo{volume}{4}},
  \bibinfo{pages}{186--197} (\bibinfo{year}{2008}).

\bibitem{Tomita2015}
\bibinfo{author}{Tomita, T.}, \bibinfo{author}{Kuga, K.},
  \bibinfo{author}{Uwatoko, Y.}, \bibinfo{author}{Coleman, P.} \&
  \bibinfo{author}{Nakatsuji, S.}
\newblock \bibinfo{title}{Strange metal without magnetic criticality}.
\newblock \emph{\bibinfo{journal}{Science}} \textbf{\bibinfo{volume}{349}},
  \bibinfo{pages}{506--509} (\bibinfo{year}{2015}).

\bibitem{Kuga2008}
\bibinfo{author}{Kuga, K.}, \bibinfo{author}{Karaki, Y.},
  \bibinfo{author}{Matsumoto, Y.}, \bibinfo{author}{Machida, Y.} \&
  \bibinfo{author}{Nakatsuji, S.}
\newblock \bibinfo{title}{Superconducting properties of the non-{Fermi}-liquid
  system $\ensuremath{\beta}\mathrm{\text{-}}\mathrm{YbAlB}_{4}$}.
\newblock \emph{\bibinfo{journal}{Phys. Rev. Lett.}}
  \textbf{\bibinfo{volume}{101}}, \bibinfo{pages}{137004}
  (\bibinfo{year}{2008}).

\bibitem{Kuga2012}
\bibinfo{author}{Kuga, K.}, \bibinfo{author}{Morrison, G.},
  \bibinfo{author}{Treadwell, L.}, \bibinfo{author}{Chan, J.~Y.} \&
  \bibinfo{author}{Nakatsuji, S.}
\newblock \bibinfo{title}{Magnetic order induced by {Fe} substitution of {Al}
  site in the heavy-fermion systems
  $\ensuremath{\alpha}\mathrm{\text{-}}\mathrm{YbAlB}_{4}$ and
  $\ensuremath{\beta}\mathrm{\text{-}}\mathrm{YbAlB}_{4}$}.
\newblock \emph{\bibinfo{journal}{Phys. Rev. B}} \textbf{\bibinfo{volume}{86}},
  \bibinfo{pages}{224413} (\bibinfo{year}{2012}).

\bibitem{Kuga2018}
\bibinfo{author}{Kuga, K.} \emph{et~al.}
\newblock \bibinfo{title}{Quantum valence criticality in a correlated metal}.
\newblock \emph{\bibinfo{journal}{Science Advances}}
  \textbf{\bibinfo{volume}{4}}, \bibinfo{pages}{eaao3547}
  (\bibinfo{year}{2018}).

\bibitem{Gegenwart2016}
\bibinfo{author}{Gegenwart, P.}
\newblock \bibinfo{title}{Gr\"uneisen parameter studies on heavy fermion
  quantum criticality}.
\newblock \emph{\bibinfo{journal}{Report on Progress in Physics}}
  \textbf{\bibinfo{volume}{79}}, \bibinfo{pages}{114502}
  (\bibinfo{year}{2016}).

\bibitem{Gegenwart2006}
\bibinfo{author}{Gegenwart, P.}, \bibinfo{author}{Weickert, F.},
  \bibinfo{author}{Garst, M.}, \bibinfo{author}{Perry, R.~S.} \&
  \bibinfo{author}{Maeno, Y.}
\newblock \bibinfo{title}{Metamagnetic quantum criticality in
  $\mathrm{Sr}_{3}\mathrm{Ru}_{2}\mathrm{O}_{7}$ studied by thermal expansion}.
\newblock \emph{\bibinfo{journal}{Phys. Rev. Lett.}}
  \textbf{\bibinfo{volume}{96}}, \bibinfo{pages}{136402}
  (\bibinfo{year}{2006}).

\bibitem{Weickert2010}
\bibinfo{author}{Weickert, F.}, \bibinfo{author}{Brando, M.},
  \bibinfo{author}{Steglich, F.}, \bibinfo{author}{Gegenwart, P.} \&
  \bibinfo{author}{Garst, M.}
\newblock \bibinfo{title}{Universal signatures of the metamagnetic quantum
  critical endpoint: Application to $\mathrm{CeRu}_{2}\mathrm{Si}_{2}$}.
\newblock \emph{\bibinfo{journal}{Phys. Rev. B}} \textbf{\bibinfo{volume}{81}},
  \bibinfo{pages}{134438} (\bibinfo{year}{2010}).

\bibitem{Kitagawa2011}
\bibinfo{author}{Kitagawa, S.} \emph{et~al.}
\newblock \bibinfo{title}{Metamagnetic behavior and kondo breakdown in
  heavy-fermion $\mathrm{CeFePO}$}.
\newblock \emph{\bibinfo{journal}{Phys. Rev. Lett.}}
  \textbf{\bibinfo{volume}{107}}, \bibinfo{pages}{277002}
  (\bibinfo{year}{2011}).

\bibitem{Kohori2001}
\bibinfo{author}{Kohori, Y.} \emph{et~al.}
\newblock \bibinfo{title}{{NMR} and {NQR} studies of the heavy fermion
  superconductors {CeTIn$_5$} ({T=Co} and {Ir})}.
\newblock \emph{\bibinfo{journal}{Phys. Rev. B}} \textbf{\bibinfo{volume}{64}},
  \bibinfo{pages}{134526} (\bibinfo{year}{2001}).
\newblock \urlprefix\url{https://link.aps.org/doi/10.1103/PhysRevB.64.134526}.

\bibitem{Kitagawa2013}
\bibinfo{author}{Kitagawa, S.}, \bibinfo{author}{Ishida, K.},
  \bibinfo{author}{Nakamura, T.}, \bibinfo{author}{Matoba, M.} \&
  \bibinfo{author}{Kamihara}.
\newblock \bibinfo{title}{Ferromagnetic quantum critical point induced by
  tuning the magnetic dimensionality of the heavy-fermion iron oxypnictide
  {Ce}({Ru}{$_{1-x}$}{Fe}{$_{x}$}){P}{O}}.
\newblock \emph{\bibinfo{journal}{Journal of the Physical Society of Japan}}
  \textbf{\bibinfo{volume}{82}}, \bibinfo{pages}{033704}
  (\bibinfo{year}{2013}).

\bibitem{Loehneysen2007}
\bibinfo{author}{L\"ohneysen, H.~v.}, \bibinfo{author}{Rosch, A.},
  \bibinfo{author}{Vojta, M.} \& \bibinfo{author}{W\"olfle, P.}
\newblock \bibinfo{title}{Fermi-liquid instabilities at magnetic quantum phase
  transitions}.
\newblock \emph{\bibinfo{journal}{Rev. Mod. Phys.}}
  \textbf{\bibinfo{volume}{79}}, \bibinfo{pages}{1015--1075}
  (\bibinfo{year}{2007}).

\bibitem{Holanda2011}
\bibinfo{author}{Holanda, L.~M.} \emph{et~al.}
\newblock \bibinfo{title}{Quantum critical {Kondo} quasiparticles probed by
  {E}{S}{R} in $\ensuremath{\beta}\mathrm{\text{-}}\mathrm{YbAlB}_{4}$}.
\newblock \emph{\bibinfo{journal}{Phys. Rev. Lett.}}
  \textbf{\bibinfo{volume}{107}}, \bibinfo{pages}{026402}
  (\bibinfo{year}{2011}).

\bibitem{Nevidomskyy2009}
\bibinfo{author}{Nevidomskyy, A.~H.} \& \bibinfo{author}{Coleman, P.}
\newblock \bibinfo{title}{Layered {Kondo} lattice model for quantum critical
  $\ensuremath{\beta}\mathrm{\text{-}}\mathrm{YbAlB}_{4}$}.
\newblock \emph{\bibinfo{journal}{Phys. Rev. Lett.}}
  \textbf{\bibinfo{volume}{102}}, \bibinfo{pages}{077202}
  (\bibinfo{year}{2009}).

\bibitem{Lawrence1981}
\bibinfo{author}{Lawrence, J.~M.}, \bibinfo{author}{Riseborough, P.~S.} \&
  \bibinfo{author}{Parks, R.~D.}
\newblock \bibinfo{title}{Valence fluctuation phenomena}.
\newblock \emph{\bibinfo{journal}{Reports on Progress in Physics}}
  \textbf{\bibinfo{volume}{44}}, \bibinfo{pages}{1} (\bibinfo{year}{1981}).

\bibitem{Bareille2018}
\bibinfo{author}{Bareille, C.} \emph{et~al.}
\newblock \bibinfo{title}{Kondo hybridization and quantum criticality in
  {$\beta$-YbAlB$_4$} by laser-{ARPES}}.
\newblock \emph{\bibinfo{journal}{Phys. Rev. B}} \textbf{\bibinfo{volume}{92}},
  \bibinfo{pages}{045112} (\bibinfo{year}{2018}).

\bibitem{Eoin2009}
\bibinfo{author}{O'Farrell, E. C.~T.} \emph{et~al.}
\newblock \bibinfo{title}{Role of $f$ electrons in the {Fermi} surface of the
  heavy fermion superconductor
  $\ensuremath{\beta}\mathrm{\text{-}}\mathrm{YbAlB}_{4}$}.
\newblock \emph{\bibinfo{journal}{Phys. Rev. Lett.}}
  \textbf{\bibinfo{volume}{102}}, \bibinfo{pages}{216402}
  (\bibinfo{year}{2009}).

\bibitem{Eoin2012}
\bibinfo{author}{O'Farrell, E. C.~T.}, \bibinfo{author}{Matsumoto, Y.} \&
  \bibinfo{author}{Nakatsuji, S.}
\newblock \bibinfo{title}{Evolution of $c\mathrm{\text{-}}f$ hybridization and
  two-component {Hall} effect in
  $\ensuremath{\beta}\mathrm{\text{-}}\mathrm{YbAlB}_{4}$}.
\newblock \emph{\bibinfo{journal}{Phys. Rev. Lett.}}
  \textbf{\bibinfo{volume}{109}}, \bibinfo{pages}{176405}
  (\bibinfo{year}{2012}).

\bibitem{EoinThesis}
\bibinfo{author}{O'Farrell, E. C.~T.}
\newblock \emph{\bibinfo{title}{Experimental studies of magnetic field tuned
  quantum criticality in {$\beta$-YbAlB$_4$} and {Sr$_3$Ru$_2$O$_7$}, PhD
  Thesis}} (\bibinfo{year}{2010}).

\bibitem{Watanabe2010}
\bibinfo{author}{Watanabe, S.} \& \bibinfo{author}{Miyake, K.}
\newblock \bibinfo{title}{Quantum valence criticality as an origin of
  unconventional critical phenomena}.
\newblock \emph{\bibinfo{journal}{Phys. Rev. Lett.}}
  \textbf{\bibinfo{volume}{105}}, \bibinfo{pages}{186403}
  (\bibinfo{year}{2010}).
\newblock
  \urlprefix\url{https://link.aps.org/doi/10.1103/PhysRevLett.105.186403}.

\bibitem{Kobayashi2018}
\bibinfo{author}{Oura, M.} \emph{et~al.}
\newblock \bibinfo{title}{Valence fluctuating compound {$\alpha$-YbAlB$_4$}
  studied by {$^{174}$Yb} {M\"ossbauer} spectroscopy and {X}-ray diffraction
  using synchrotron radiation}.
\newblock \emph{\bibinfo{journal}{Physica B: Condensed Matter}}
  \textbf{\bibinfo{volume}{536}}, \bibinfo{pages}{162 -- 164}
  (\bibinfo{year}{2018}).
\newblock
  \urlprefix\url{http://www.sciencedirect.com/science/article/pii/S092145261730621X}.

\bibitem{Tokiwa2015}
\bibinfo{author}{Tokiwa, Y.}, \bibinfo{author}{Stingl, C.},
  \bibinfo{author}{Kim, M.-S.}, \bibinfo{author}{Takabatake, T.} \&
  \bibinfo{author}{Gegenwart, P.}
\newblock \bibinfo{title}{Characteristic signatures of quantum criticality
  driven by geometrical frustration}.
\newblock \emph{\bibinfo{journal}{Science Advances}}
  \textbf{\bibinfo{volume}{1}}, \bibinfo{pages}{e1500001}
  (\bibinfo{year}{2015}).

\bibitem{MatsumotoScience}
\bibinfo{author}{Matsumoto, Y.} \emph{et~al.}
\newblock \bibinfo{title}{Quantum criticality without tuning in the mixed
  valence compound $\ensuremath{\beta}$-$\mathrm{YbAlB}_{4}$}.
\newblock \emph{\bibinfo{journal}{Science}} \textbf{\bibinfo{volume}{331}},
  \bibinfo{pages}{316--319} (\bibinfo{year}{2011}).

\bibitem{Dzero2010}
\bibinfo{author}{Dzero, M.}, \bibinfo{author}{Sun, K.},
  \bibinfo{author}{Coleman, P.} \& \bibinfo{author}{Galitski, V.}
\newblock \bibinfo{title}{Topological {K}ondo insulators}.
\newblock \emph{\bibinfo{journal}{Phys. Rev. Lett.}}
  \textbf{\bibinfo{volume}{104}}, \bibinfo{pages}{106408}
  (\bibinfo{year}{2010}).

\bibitem{Komijani2018}
\bibinfo{author}{Komijani, Y.} \& \bibinfo{author}{Coleman, P.}
\newblock \bibinfo{title}{Model for ferromagnetic quantum critical point in a
  1{D} {Kondo} lattice}.
\newblock \emph{\bibinfo{journal}{Phys. Rev. Lett.}}
  \textbf{\bibinfo{volume}{120}}, \bibinfo{pages}{157206}
  (\bibinfo{year}{2018}).

\bibitem{Macaluso}
\bibinfo{author}{Macaluso, R.~T.} \emph{et~al.}
\newblock \bibinfo{title}{Crystal structure and physical properties of
  polymorphs of {LnAlB$_4$} ({Ln = Yb, Lu})}.
\newblock \emph{\bibinfo{journal}{Chemistry of Materials}}
  \textbf{\bibinfo{volume}{19}}, \bibinfo{pages}{1918--1922}
  (\bibinfo{year}{2007}).
\newblock \urlprefix\url{https://doi.org/10.1021/cm062244+}.
\newblock \eprint{https://doi.org/10.1021/cm062244+}.

\bibitem{MagnetostrictionRSI}
\bibinfo{author}{K\"{u}chler, R.}, \bibinfo{author}{Bauer, T.},
  \bibinfo{author}{Brando, M.} \& \bibinfo{author}{Frank, S.}
\newblock \bibinfo{title}{A compact and miniaturized high resolution
  capacitance dilatometer for measuring thermal expansion and
  magnetostriction}.
\newblock \emph{\bibinfo{journal}{Review of Scientific Instruments}}
  \textbf{\bibinfo{volume}{83}}, \bibinfo{pages}{095102}
  (\bibinfo{year}{2012}).

\bibitem{CalorimetryRSI}
\bibinfo{author}{Matsumoto, Y.} \& \bibinfo{author}{Nakatsuji, S.}
\newblock \bibinfo{title}{Relaxation calorimetry at very low temperatures for
  systems with internal relaxation}.
\newblock \emph{\bibinfo{journal}{Review of Scientific Instruments}}
  \textbf{\bibinfo{volume}{89}}, \bibinfo{pages}{033908}
  (\bibinfo{year}{2018}).

\bibitem{MagnetizationM}
\bibinfo{author}{Sakakibara, T.}, \bibinfo{author}{Mitamura, H.},
  \bibinfo{author}{Tayama, T.} \& \bibinfo{author}{Amitsuka, H.}
\newblock \bibinfo{title}{Faraday force magnetometer for high-sensitivity
  magnetization measurements at very low temperatures and high fields}.
\newblock \emph{\bibinfo{journal}{Japanese Journal of Applied Physics}}
  \textbf{\bibinfo{volume}{33}}, \bibinfo{pages}{5067} (\bibinfo{year}{1994}).
\newblock \urlprefix\url{http://stacks.iop.org/1347-4065/33/i=9R/a=5067}.

\bibitem{OFarrell_IOP_2012}
\bibinfo{author}{O'Farrell, E. C.~T.}, \bibinfo{author}{Tompsett, D.~A.},
  \bibinfo{author}{Horie, N.}, \bibinfo{author}{Nakatsuji, S.} \&
  \bibinfo{author}{Sutherland, M.~L.}
\newblock \bibinfo{title}{Shubnikov-de {Haas} oscillations in the heavy fermion
  $\alpha$-{YbAlB}$_4$}.
\newblock \emph{\bibinfo{journal}{Journal of Physics: Conference Series}}
  \textbf{\bibinfo{volume}{391}}, \bibinfo{pages}{012053}
  (\bibinfo{year}{2012}).
\newblock
  \urlprefix\url{https://doi.org/10.1088%2F1742-6596%2F391%2F1%2F012053}.

\bibitem{Takano2016}
\bibinfo{author}{Takano, S.} \emph{et~al.}
\newblock \bibinfo{title}{Site-selective $^{11}${B} {NMR} studies on
  {YbAlB}$_4$}.
\newblock \emph{\bibinfo{journal}{Journal of Physics: Conference Series}}
  \textbf{\bibinfo{volume}{683}}, \bibinfo{pages}{012008}
  (\bibinfo{year}{2016}).

\end{thebibliography}
\bibliographystyle{naturemag} 

\vspace{1 cm} \noindent{\bf Acknowledgments} We acknowledge the help of Naoki Horie. M.S.G. acknowledges the support of Croatian Science Foundation (HRZZ) under the project IP-2018-01-2970, the Unity Through Knowledge Fund (UKF Grant No. 20/15) and the support of project CeNIKS co-financed by the Croatian Government and the European Union through the European Regional Development Fund - Competitiveness and Cohesion Operational Programme (Grant No. KK.01.1.1.02.0013). The work at Rice University (A.H.N.) was supported by the U.S. National Science Foundation CAREER grant no. DMR-1350237 and the Welch Foundation grant C-1818. R.K. is supported by the German Science Foundation through Project No. KU 3287/1-1. M. L. S. acknowledges the support of the EPSRC and the Royal Society.
 
\vspace{0.5 cm}
 

\noindent{\bf Author contributions} E.C.T.O., M.S.G. and Y.M. contributed equally to this work. E.C.T.O. performed the  resistivity, Hall effect and quantum oscillations measurements, M.S.G., M.Y. and M.T. performed NMR experiments, Y.M. performed specific heat measurements. Y.M., K.K., Y.K., Y.S. and T.S. performed magnetization measurements and K.K. synthesized the samples. R.K. and Y.M. performed the thermal expansion and magnetostriction measurements. M.L.S., M.B. and A.H.N. contributed to data interpretation. M.T. and S.N. conceived the project, planed the research and contributed to data interpretation. E.C.T.O., M.S.G. and Y.M. wrote the paper. All authors took part in discussing results and editing the manuscript.

\vspace{0.5cm} \noindent{\bf Correspondence} and requests for materials should be addressed to E. C. T. O., M.S.G, Y. M., M.T. and S.N.

\vspace{0.5cm} \noindent{\bf Competing financial interests} The authors declare no competing
financial interests. 

\end{document}